\documentclass[11pt, times]{article}

\usepackage{geometry}                
\usepackage{amssymb}
\usepackage{placeins}
\usepackage[numbers, sort&compress]{natbib}
\usepackage{varioref}	
\usepackage[colorlinks, citecolor=blue]{hyperref} 	
\usepackage{hypcap} 	
\usepackage{gensymb} 	
\usepackage{textcomp}	
\usepackage{enumerate}	
\usepackage[breakable]{tcolorbox} 
\usepackage{amsmath}
\usepackage{enumitem}
\usepackage{booktabs} 

\title{Influence of drying conditions on the stress and weight development of capillary suspensions}

\author{Steffen B. Fischer$^\mathrm{a,b}$, Erin Koos$^\mathrm{a,\ast}$ }

\date{\small
    $^\mathrm{a}$KU Leuven, Soft Matter, Rheology and Technology, Department of Chemical Engineering, Celestijnenlaan 200f, 3001 Leuven, Belgium \\
    $^\mathrm{b}$Karlsruhe Institute of Technology, Institute for Mechanical Process Engineering and Mechanics, Karlsruhe, Germany 
    ~\\
    ~\\
    $^{\ast}$ e-mail: erin.koos@kuleuven.be \\
}

\begin{document}
\maketitle

\begin{abstract}
	\centerline{\includegraphics[width=\linewidth]{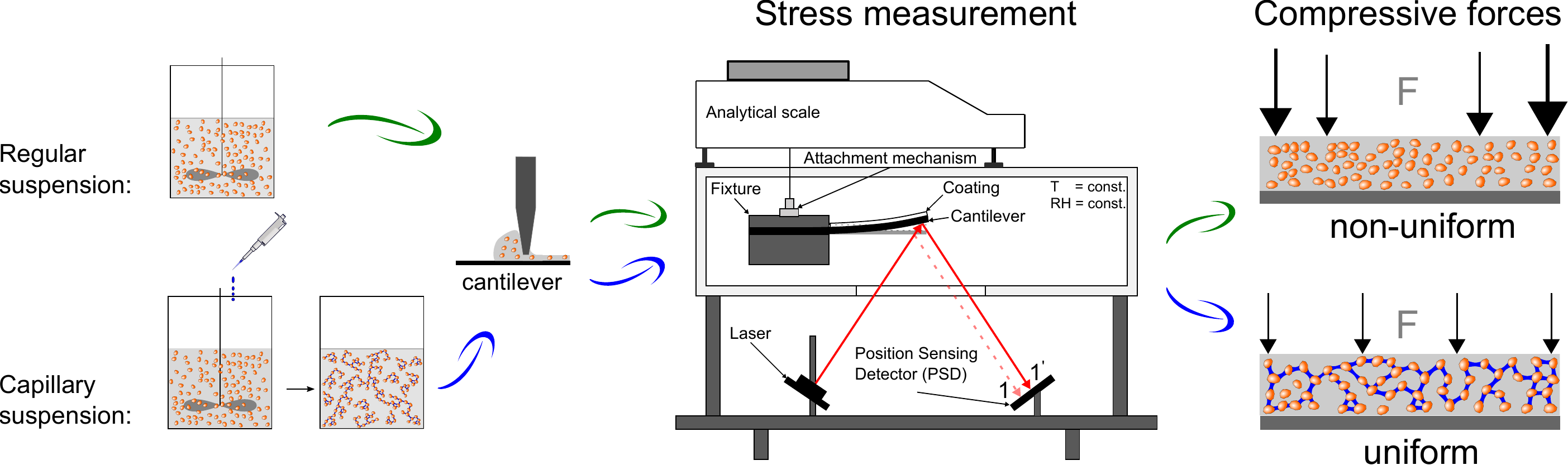}}
	\label{fig:graphical-abstractuniform-drying}
Cracking of suspensions during drying is a common problem. While additives, e.g. binders and surfactants, can mitigate this problem, some applications, such as printing conductive pastes or sintering green bodies, do not lend themselves to the use of additives. Capillary suspensions provide an alternative formulation without additives. In this work, we use simultaneous stress and weight measurements to investigate the influence of formulation and drying conditions. Capillary suspensions dry more homogeneously and with lower peak stresses, leading to an increased robustness against cracking compared. An increase in dry film porosity is not the key driver for the stress reduction. Instead, the capillary bridges, which create strong particle networks, resist the stress. Increasing the relative humidity enhances this effect, even for pure suspensions. While lower boiling point secondary liquids, e.g. water, persist for very long times during drying, higher boiling point liquids offer further potential to tune the the drying process. 


\end{abstract}

\section{Introduction}
\label{Intro}

Stress growth during drying of paints, inks and coatings is an important measure for developing defect free coatings. Stress is a direct measure of a film's proneness to cracking. In nature, we find crack patterns in things like mud cracks \cite{Goehring.2010}. Examples of technical importance include drying of ceramic tape cast films \cite{Hotza.1995}, as well as subsequent green body binder burn out, where internal stresses can build up and lead to cracking \cite{Tsai.1991, BohnleinMau.1992, Fu.2015b}. Another example is represented by screen printing of conductive inks \cite{Kamyshny.2014}. With increasing demand for low-cost solar cells and electronic gadgets, such as RFID tags, high throughput production is required \cite{Habas.2010}. The functional material is printed on flexible polymeric substrates with low glass transition temperatures. If drying of these electrically conducting circuits lead to cracks, their function would be destroyed. In order to print the circuits, or apply pigments, the functional particles have to be suspended in a liquid, forming a suspension. Upon application, drying sets in. Initially, the suspension dries continuously as if there were no particles present for as long as the surface is covered with liquid. This drying behavior is termed the constant rate period (CRP) \cite{Scherer.1990b}. Due to the evaporation of the liquid, the film shrinks and the solids density (particle volume fraction) increases. Generally, drying does not occur uniformly across the coating on narrow substrates or in droplet evaporation. In dilute systems, such as drying of a coffee droplet, this leads to the coffee-ring effect in which particles are transported towards the edges. The cause for this phenomenon, capillary flows, was identified by Deegan et al.~\cite{Deegan.1997}. As the droplet starts shrinking, the three phase contact line remains pinned at the substrate. Since the droplet perimeter stays constant and evaporation continues, there must be flow carrying particles towards the edges forming stains. The same phenomenon occurs during the drying of more concentrated suspensions. As the film dries laterally, transporting particles to the edges, particle depleted areas, or film defects,  such as pinholes or trenches are left behind \cite{Chiu.1993b, Guo.1999, Holmes.2008, Holmes.2006, Ma.2005, Routh.1998, Tirumkudulu.2004, Fischer.2020}. When drying proceeds and a compact film forms, the CRP decreases due to imposed mass transport resistances. The liquid filled voids in the saturated and consolidated particle coating represent pores and necks. Further evaporation pins the surface liquid to the pore mouths, such that concave menisci start to develop. This menisci formation causes pressure differences across the interface, which is described by the Young-Laplace equation for the capillary pressure $ p_c $:
        \begin{equation}
                \label{eq:p_cap}
                p_c= \frac{2\gamma_{lv} \cos{\theta}}{r}
        \end{equation}
Where the capillary pressure $ p_\mathrm{c} $, depends on the liquid-vapor interfacial tension $ \gamma_\mathrm{lv} $, the contact angle $ \theta $, and the radius of a capillary tube inscribing the neck between particles $ r $. 
Continued evaporation and pinning of the liquid causes the contact angle to decrease, which leads to a larger capillary pressure. This, in turn, compacts the surrounding particles. If the coating is free to shrink in all directions, no stresses are observed. However, if the coating adheres to the substrate, which locally restricts the shrinkage, the strain mismatch can lead to severe cracking in hard particles or film formation for soft particles \cite{Lei.2001, Lei.2002, Singh.2007}. In order to prevent film fracture, there are different possibilities to modify the suspension. Avoiding the formation of small pores and particle mobility is achieved by addition of binders, which store the stresses and prevent delamination between the film and substrate. 

Additionally, the flow properties of the suspension need to be adjusted for the application method. In many coatings, carboxymethyl cellulose (CMC) is used as thickening agent. However, a study by Wedin et al. showed a dramatic increase in peak stress during drying of a coating when CMC was added \cite{Wedin.2004}. Furthermore, \autoref{eq:p_cap} shows the direct relationship between the capillary pressure and liquid-vapor interfacial tension, which can be lowered through the addition of surfactants. In order to obtain an electrically conductive circuit, these additives have to be removed, which is usually done by heat treatment. However, on flexible, polymeric substrates with low glass transition temperatures, the treatment is restricted in terms of temperature. An alternative to solid additives that need additional treatment is provided by capillary suspensions. In capillary suspensions, the ``binder" is a second liquid, immiscible with the bulk phase \cite{Koos.2011}. Upon the addition of a very small volume percent of this immiscible liquid, a dramatic change in rheological properties is observed \cite{Koos.2014}. The obtained capillary suspension exhibits an increase in the yield stress by several orders of magnitude, while showing shear-thinning behavior averting the need for potentially stress-inducing CMC, making these suspensions ideally suited for various printing applications \cite{Maurath.2017, Schneider.2016, Yuce.2018} or ceramic bodies \cite{Wei.2019}. This change in properties is caused by the sample spanning particle network induced by capillary bridges of the secondary liquid. In our previous work, we found these novel suspensions reduce cracking without addition of further additives \cite{Schneider.2017}. More recently, we found capillary suspensions form a uniform final coating after drying, devoid of pinholes and trenches despite initially present lateral drying \cite{Fischer.2020}. 

Since cracking is driven by stress build-up, stress measurements indicate the resilience of a coating. Many previous studies have tracked the stress formation of coatings using the cantilever deflection method due to its simple concept \cite{Payne.1997, Kim.2009, Lei.2001, Price.2015, Tirumkudulu.2004, Routh.2013}. Moreover, this method allows the cantilever to be placed in a controlled environmental chamber. Besides the temporal stress measurement, the change in mass over time, i.e. drying rate, is another important factor in the coating's drying behavior. However, this largely complicates the experimental setup. In several studies, the drying rate was obtained by coating another substrate, which was then weighed inside the same chamber, while the stress was tracked on the other substrate \cite{Francis.2002, Martinez.2002b, Wedin.2004}. This approach works well, as long as the drying conditions in the chamber are spatially homogeneous and, most importantly, the coatings are identical. For capillary suspensions with their high yield stress, the sample application on the cantilever has proven to be challenging without introducing misinterpretations. Ideally, the stress development and weight loss are measured simultaneously. Studies by Kiennemann et al.~\cite{Kiennemann.2005} and Fu et al.~\cite{Fu.2015} overcame these limitations by placing the clamped cantilever directly on an analytical balance. However, since the balance cannot be placed inside a small chamber with uniform drying conditions, their experiments were carried out under ambient conditions that are difficult to control. Our apparatus design, allows the simultaneous tracking of weight loss and stress development in the same coating inside a drying chamber under controlled conditions. This paper examines the differences in these quantities between a pure suspension and a capillary suspension with different types and amounts of secondary liquid under various drying conditions.

\section{Materials and Methods}

\subsection{Sample preparation}

Alumina suspensions were prepared at different particle volume fractions and with a variation of secondary liquid to obtain different rheological properties. Alumina particles ($\alpha$-Al$_2$O$_3$, CT3000SG, Almatis GmbH, Germany) with an average particle size of $ \mathrm{d_{50,3}=0.5} $~\micro m according to the supplier were dried in an oven at 100 \degree C overnight and dispersed in 1-heptanol ($ > $99\%, Alfa Aesar). Mixing of particles was performed in multiple steps (at least twice for two minutes at 3500 rpm) with a Speedmixer DAC 150.1~FV (Hauschild \& Co. KG, Germany) and in 25 ml polypropylene cups ($\varnothing$= 35 mm) to obtain smooth samples with particle volume fractions of $ \mathrm{\phi_{solid}=0.2}$ and $ \mathrm{\phi_{solid}=0.25}$. In order to create capillary suspensions, ultra-pure water (atrium 611 DI, Sartorius AG, Germany) was added with a micropipette to achieve the desired water volume fractions of $\mathrm{\phi_{H_2O}}$. These fractions were chosen to keep the ratio of the volume of the secondary liquid bridges ($ \mathrm{\phi_{sec}} $) to the particle volume fraction ($ \mathrm{\phi_{solid}} $) constant, as shown in \autoref{tab:coating}. 
        \begin{table*}[]
            	\centering
            	\caption{Composition and coating parameters for each sample tested.} 
            	\begin{tabular}{c c c c c } \toprule
            		$\mathrm{\phi_{solid}}$
            			& sec. fluid	& $\mathrm{\phi_{sec}/\phi_{solid}} $	
            								& Gap height [\micro m] 
            										& Coating speed [m/s] \\ \midrule
            		0.20	&  -- 			& 0 		& 250	& 0.07 \\
            		0.20	& water  		& 0.075 	& 250 	& 0.09 \\
            		0.20	& water  		& 0.125 	& 230 	& 0.29 \\
            		0.20	& glycerol  	& 0.125 	& 330 	& 0.17 \\ 
            		0.25	& --  			& 0 		& 320 	& 0.07 \\
            		0.25	& water  		& 0.124 	& 170 	& 0.29 \\
            		0.25	& glycerol 		& 0.125 	& 170 	& 0.17 \\ \bottomrule
            	\end{tabular}
            	\label{tab:coating}
        \end{table*}
For spectrometry measurements, we used heavy water ($ \mathrm {D_{2}O} $, 99.9 atom\% deuterium, Sigma-Aldrich). Additionally, we prepared samples with glycerol ($ > $99.0\%, GC grade, Sigma-Aldrich), at the same volume fractions. After the addition of the secondary liquid, the samples were again mixed at 3500~rpm in the Speedmixer. The duration depended on the smoothness of the samples after one mixing step of two minutes. If agglomerates were still present, another two minute mixing step was applied until agglomerates were no longer visible. The reason for stepwise mixing is to prevent the sample from heating. The samples were stored in the polypropylene Speedmixer cups wrapped in parafilm and remixed before use.

\subsection{Coating process}
\label{sec:coating}

\begin{figure}[t!]
            	\capstart 
            	\centering
            	\includegraphics[width=0.75\textwidth]{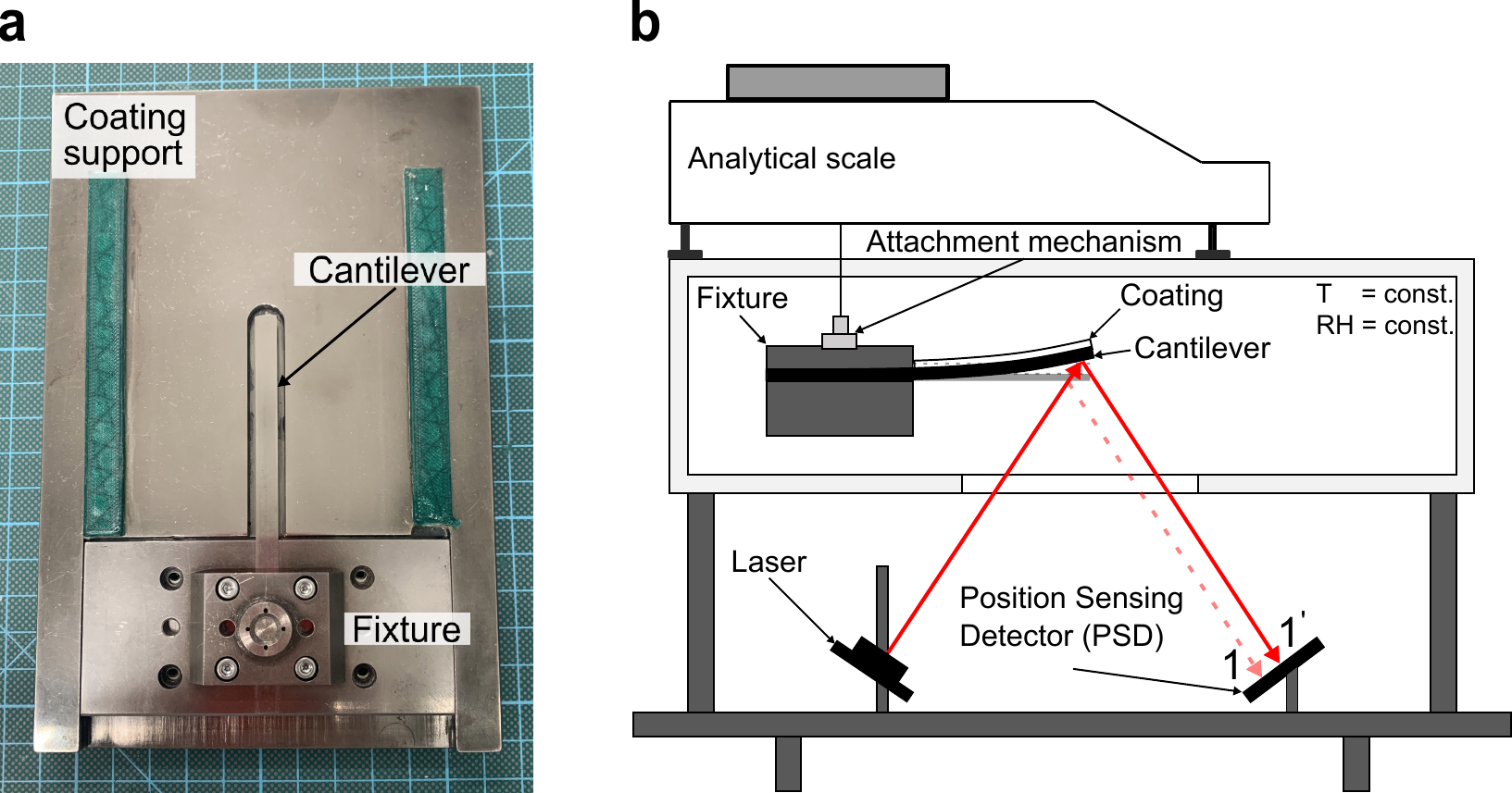}
            	\caption{(\textbf{a}) The steel coating rig with the clamped cantilever fixture placed in the recess of the rig. To maintain the cantilever's position, it is supported from below. Upon sample deposition, a coating blade with preset gap height is moved between the green rails along the cantilever with a constant coating velocity. (\textbf{b}) Schematic of the humidity and temperature controlled drying chamber. After coating, the fixture is inserted into the chamber, and attached to the analytical balance. During the measurement, the weight loss and cantilever deflection, tracked by the position sensing detector, are recorded.}
            	\label{fig:method}
        \end{figure}

In order to repeatably and uniformly coat the cantilever, we manufactured the small coating rig shown in \autoref{fig:method}a. 
At first, the stainless steel cantilever was clamped perpendicularly in the fixture between two sanded steel slabs such that the available area for the coating measures 6~mm wide and 40~mm long. The cantilever's thickness is 200~\micro m. After clamping, the fixture was then placed in the recess of the coating rig, where the cantilever was supported to prevent bending while coating. This bending can lead to an inhomogeneous wet film thickness along the cantilever. Before use, the samples were remixed at 3500~rpm for one minute, in order to have a well-mixed suspension before application. The pastes were deposited onto the cantilever and spread with a coating knife (ZUA 2000, Zehntner GmbH, Sissach, Switzerland) by means of dragging the blade between the green rails along the cantilever, driven by a voltage regulated motor. The high yield stress capillary suspension sample with ($\mathrm{\phi_{H_{2}O}=0.025} $) required a pre-coating step with a spatula to spread the sample. Besides the high network strength (yield stress) and shear thinning behavior, capillary suspensions also tend to exhibit wall-slip \cite{Koos.2014b}. These properties make it necessary to adapt the respective coating settings. For example, the pure and lower magnitude capillary suspension ($\mathrm{\phi_{H_{2}O}=0.015}$) were coated at the same gap height of 250~\micro m but different drying speeds. Both quantities are summarized in \autoref{tab:coating} for each formulation. The dry film thickness was kept constant at $ \mathrm{75 \pm 5 }$~\micro m.

\subsection{Stress measurement in the environmentally controlled drying chamber}

Our design of the stress measurement apparatus is shown in \autoref{fig:method}b. The desired drying temperature is set on the controller and allowed to equilibrate for at least one hour. The desired relative humidity is obtained by manually mixing water saturated air, which was passed through a bubbler, with dry air at a total air flow rate of $ \leq $8 liters per minute.  After coating the cantilever according to the method described in \autoref{sec:coating}, the fixture with the clamped cantilever is quickly inserted into the drying chamber. This is done by means of a specially developed sliding and lifting device. The fixture is placed on the device, pushed inside the chamber and lifted so that it is connected to the attachment mechanism suspended from the analytical balance (Sartorius Cubis MSA 224S, Sartorius Lab Instruments GmbH \& Co.KG, G\"{o}ttingen, Germany) as shown in \autoref{fig:method}b. The attachment mechanism was designed to ensure the correct alignment of the clamped cantilever and is maintained between measurements and throughout the entire drying process. The mechanism, lifter, and door also allow the measurement to be started quickly ($ \sim $30 seconds) after coating. After closing the chamber door, data acquisition is started. Laser light is directed to the polished bottom surface two mm from the free cantilever end where it is reflected and subsequently captured by the position sensing detector (PSD). 

As drying proceeds, film shrinkage confined by good adhesion of the film to the substrate causes stresses, which are captured as bending in the cantilever. This bent cantilever causes the deflected laser beam to travel on the PSD from point 1 to $ 1' $. Before the actual measurement, the setup was calibrated to correlate the movement of the laser on the PSD with the deflection of the cantilever. This was done by deliberately deflecting the cantilever a known distance with a micrometer screw while the entire fixture is in the measurement position. The stress in the deformed substrate was derived by Stoney in 1909 \cite{Stoney.1909} for an uniaxial film stress, which was later corrected to a biaxial in plane stress due to an influence of the cantilever width, which is larger than the coating height \cite{Janssen.2009}. Corcoran~\cite{Corcoran.1969} later developed a correlation that takes the effect of substrate bending on the developed coating stress into account and is now widely used for determining stresses in drying coatings. The stress $ \sigma $ within the coating is given by
        \begin{equation}
                \label{eq:Corcoran}
                \sigma = \frac{d\cdot E_s t_s^3}{3 t_c L^2 \cdot (t_s + t_c)(1- \nu_s)} + \frac{d \cdot E_c(t_s + t_c)}{L^2 \cdot (1-\nu_c)} 
        \end{equation}
where subscripts $ s $ and $ c $ represent substrate and coating, respectively. The substrate (cantilever)  thickness $ t_s = 0.196 $~mm, the Young's modulus $ E_s = 210 $~GPa, Poisson's ratio $ \nu_s = 0.3 $, and the cantilever's free length $ L = 40 $~mm.

However, the correlation contains quantities that are difficult to obtain. The coating thickness at the moment of drying $ t_c $ is unknown and instead the dry film thickness is used. Another unknown parameter is the coating's Young's modulus $ E_c $, which is difficult to obtain in general and especially at different drying times. This drawback can be overcome if the experimental design is chosen such that the second term in \autoref{eq:Corcoran} can be neglected. In a similar study using alumina particles with added binders, the measured maximum Young's modulus after drying was 8.5~GPa \cite{Fu.2015}. The stainless steel cantilever in our study has a Young's modulus of 210~GPa, which leads to the first requirement for neglecting the second term: $ E_s >> E_c $. With a dry coating thickness $ t_\mathrm{c}< 80 $ ~\micro m, the induced error by neglecting the second term is expected to be below 10\% \cite{Fu.2015}. Moreover, several assumptions underlay Corcoran's equation, such as ideal adhesion of the coating on the substrate, isotropic elastic properties of coating and substrate, a uniform biaxial stress in the coating, and small deflection, amongst others \cite{Corcoran.1969}. We did not observe an effect of coating weight loss on the deflection, but experiments at 40 \degree C and 50 \% RH required that the fixture be pre-heated before coating to avoid condensation when inserted into the chamber. Given all of the assumptions, the result of the measurements will be in the correct order of magnitude, but should be considered qualitatively and in a comparative manner within this study. Error in peak stress measurements are calculated from the standard deviation of at least five measurements. 

The dry film thickness ($ t_c $) was measured with a digital stylus micrometer indicator (ID-H530, Mitutoyo, Japan) with a precision of $\pm $ 1.5~\micro m. Since the dry films are very delicate, special care had to be taken in carefully lowering the measuring stylus on the coating to prevent excessive compression of the coating. The film thickness was measured at more than 15 locations distributed along the length and off center of the film.

\subsection{Shear yield stress measurements}
Yield stress measurements, which provide a measure for a sample's network strength, were performed with a vane geometry (FL100/6W/Q1, 6 vanes, 22 mm diameter, 16 mm height) on a MCR 702 rheometer (Anton Paar GmbH, Germany). In order to investigate the temperature dependence on the yield stress, the prepared and sealed samples were placed in a lab oven at a constant temperature of 30 \degree C and 40 \degree C, respectively, and allowed to equilibrate. Before measurement, the samples were quickly remixed for one minute in the Speedmixer to account for any particle settling that may have occurred during temperature equilibration. Subsequently, the sample still in the mixing cup is placed in a modified temperature controlled cup holder. Upon lowering the vane into the cup, the sample was allowed to rest for five minutes to relax any induced stresses. Afterwards, a stepwise stress ramp was applied until yield occurred. The resulting strain vs. stress curve was then analyzed with the tangent method to obtain the apparent yield stress of the sample. Reported errors are from the standard deviation of triplicate measurements.  

\subsection{Thermogravimetric analysis with mass spectrometry (TGA-MS)}
\label{sec:TGA}

The coating is assumed to be dry, once a weight change is no longer detected. However, depending of the medium to be dried, there is an isothermal sorption equilibrium between the liquid in the coating and the ambient air. This equilibration depends on the relative humidity (in case of water evaporation), but also mass transport resistances in the film. Depending on the liquid, it is often undesirable to have residual fluid left in a coating. In this work, we used two different instruments to determine the amount of liquid left in the film prepared by different formulations. At first, we used a thermogravimetric analyzer (TGA Q500, TA Instruments, USA). Less than 9~mg of the paste was weighed into a crucible. The amount was chosen such that the film height within the crucible was comparable to the film thickness on the cantilever. After placing the sample in the crucible, it was gently tapped on the table with tweezers to achieve equal spreading. Afterwards, it was placed in a suspended weighing pan attached to the TGA. The furnace was closed and the measurement started. The furnace and balance were always flushed with nitrogen gas at flow rates of 60~ml/min and 40~ml/min, respectively. The sequential stages followed the same principle. At first, the furnace was heated to the desired drying temperature where it was held until the sample was dry. Subsequently, the temperature was increased with a ramp of 50~\degree C/min to a temperature of 500~\degree C, where it was once again held constant. Afterwards, the residual fluid load was calculated as the difference in weight before the temperature ramp and the weight at the end of the high temperature hold time divided by the weight at the end. The conditions varied for different sample particle volume fractions and are shown in \autoref{tab:TGA}.
        \begin{table*}[]
            	\centering
            	\caption{TGA profile for measuring the residual fluid content}
            	\begin{tabular}{ c c c c c c} 
            		\toprule
            			& \multicolumn{2}{c}{Stage 1}	& Stage 2 & \multicolumn{2}{c}{Stage 3} \\ \cmidrule(l){2-3} \cmidrule(l){4-4} \cmidrule(l){5-6}
            			$ \mathrm{\phi_{solid}} $	& Temperature		& Duration & Ramp	& Temperature		& Duration \\\midrule
            		0.20	& 30 \degree C & 500 min & 50 \degree C/min & 500 \degree C & 60 min \\ 
            		0.20	& 40 \degree C & 430 min & 50 \degree C/min & 500 \degree C & 60 min \\
            		0.25	& 30 \degree C & 220 min & 50 \degree C/min & 500 \degree C & 60 min \\
            		0.25	& 40 \degree C & 160 min & 50 \degree C/min & 500 \degree C & 60 min \\ \bottomrule
            	\end{tabular}
            	\label{tab:TGA}
        \end{table*}

Later, we analyzed the headspace composition during drying of the capillary suspension samples at drying temperatures of 30~\degree C and 40~\degree C. In this experiment, we are interested in the sequence in which the components evaporate. In order to save equipment time, we reduced the sample weight, i.e. film thickness. As before, the samples were deposited in the crucible and gently tapped on the table for spreading. Afterwards, they were analyzed in a TGA coupled with mass spectrometry (MS). Compared to the previous instrument, this TGA (STA 449 F3 Jupiter, NETZSCH-Ger\"{a}tebau GmbH, Selb, Germany) differed in the measuring system, where the crucible was placed on a carrier system rather than being suspended. The drying gas was a mixture of 80~ml/min nitrogen and 20~ml/min oxygen to mimic air. In this measurement, only the samples with $ \mathrm{\phi_{solid}= 0.2} $ and $ \mathrm{\phi_{H_2O}/\phi_{solid}=0.125} $ were analyzed. The stage 1 profile used a shorter holding time of 100 minutes and 180 minutes for 30~\degree C and 40~\degree C, respectively. The temperature ramp in stage 2 was also lowered to 20~\degree C/min. 

Following evaporation, the gas passes through the mass spectrometer (HPR-20 QIC, Hiden Analytical Ltd., UK) by means of the carrier gas mixture. The ionized and accelerated molecules are detected by a multiple ion detection (MID) scan. The characteristic mass to charge (m/z) ratios were detected at 20 m/z for D$_2$O and 56~m/z for 1-heptanol. We chose heavy water because 1-heptanol has a fragment peak overlapping with normal water at 18~m/z. After closing the furnace, the instrument was purged for several minutes. The measurement was subsequently started and sampling on both instruments were triggered. The MS has a delay of a few minutes compared to the weight change in the balance. 

\section{Results and Discussion}
\subsection{Stress development differences in capillary suspensions}
\label{sec:stress}

\begin{figure*}[t!]
            	\capstart
            	\centering
            	\includegraphics[width=1\linewidth]{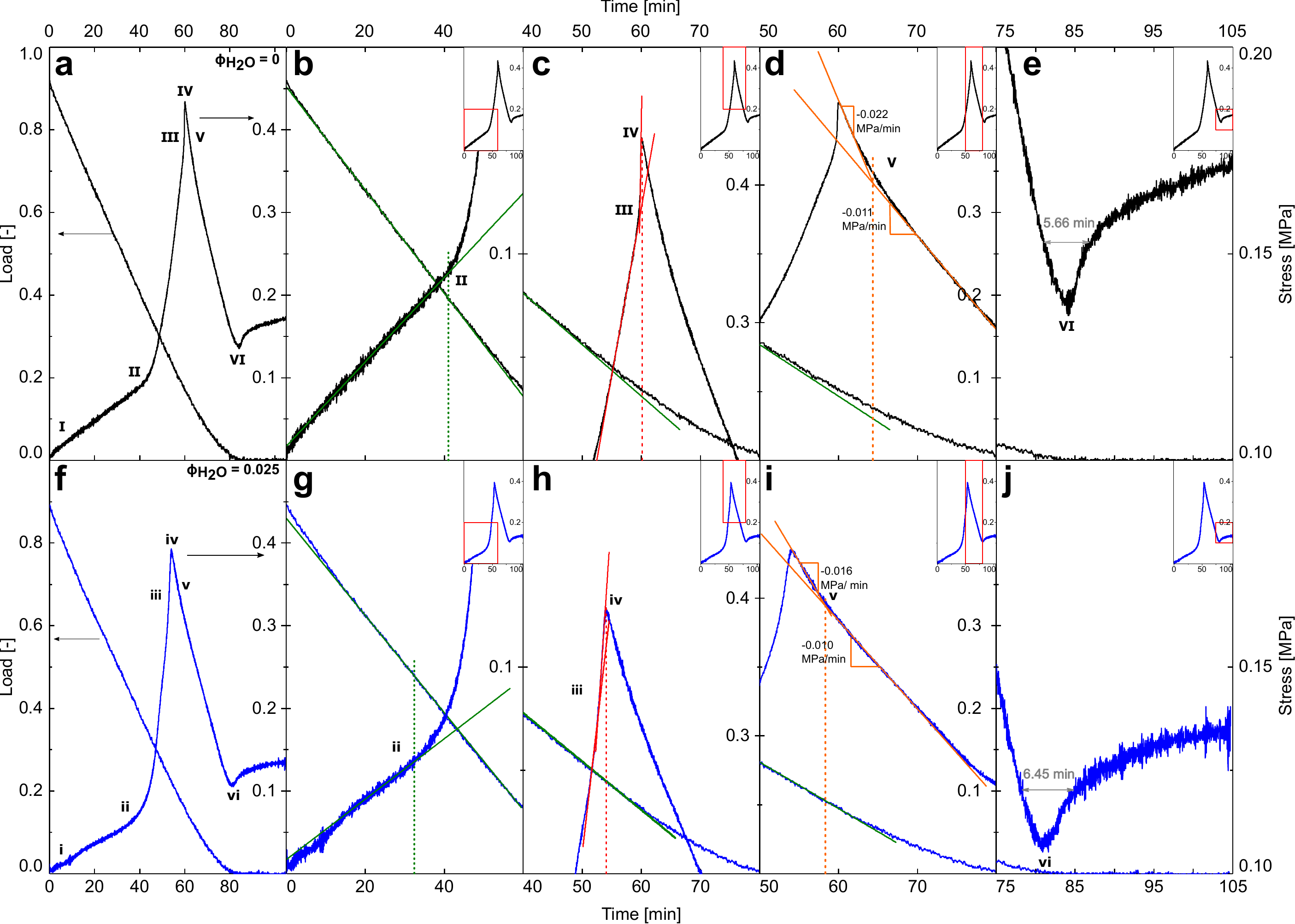}
            	\caption{The results for simultaneous stress and weight measurement at a drying temperature of 40 \degree C and 1\% RH for (\textbf{a-e}) a pure suspension sample and (\textbf{f-j}) a capillary suspension sample with 2.5 vol\% added water, each at an initial solid volume fraction of 0.2. The left most panel (a,f) shows the entire stress and load evolution over time. The magnitude for the stress value can be found on the right y-axis and with its span being different in each column but the same for both panel rows. The magnitude of the load is shown on the left y-axis in each panel, where it is shifted relative to the stress curve in each column for better visibility of trends. The load scale is the value shown on the left of each panel (the stress scaling from the previous panel). Special points of interest are marked in each panel with capitalized roman numerals for the pure suspension and lower case roman numerals for the capillary suspension.}
            	\label{fig:stress_dev}
        \end{figure*}

In \autoref{fig:stress_dev}, 
the stress and weight development comparison during drying at 40 \degree C and a relative humidity of 1\% is shown for two representative samples. The pure suspension of alumina dispersed in 1-heptanol is depicted in the upper row (\autoref{fig:stress_dev}(a-e)), while \autoref{fig:stress_dev}(f-j) shows the $ \mathrm{\phi_{solid}=0.2} $ capillary suspension sample with 2.5 vol\% water added ($ \mathrm{\phi_{H_2O}/\phi_{solid}=0.125} $). The stress value is shown on the right y-axis and the load is shown on the left y-axis. The load is defined as
        \begin{equation}
                \label{eq:load}
                \mathrm{load}=\frac{m(t)-m_\mathrm{final}}{m_\mathrm{final}}=\frac{m_\mathrm{liquid}(t)}{m_\mathrm{dry}}
        \end{equation}
with $ m(t) $, being the film mass measured at time $ t $. The measured dry film thickness for the pure and the capillary suspension were $ 71\pm1 $ \micro m and $ 69\pm4 $ \micro m, respectively. The leftmost panels (a and f) show the entire stress and load curve with excerpts thereof shown in the other panels. While the x- and y-scales differ for each of the excerpts, they are identical between the two samples (columns). The inset in each panel represents the entire stress curve with a box marking the magnified area depicted in the main panel. We have identified six key points, marked in the stress graphs with capital roman numerals for the pure suspension and lower case numerals (i-vi) for the capillary suspensions according to the following observations: 
\begin{enumerate}[label=\Roman*]
	\item start of the measurement;
	\item the point where the stress deviates from an initial linear trend;
	\item the rapid increase in stress preceding the stress peak;
	\item peak stress;
	\item end of rapid stress decrease following the stress peak;
	\item stress trough before the residual stress value is reached.
\end{enumerate}

First, we will examine the findings of the pure suspension without added water ($\mathrm{\phi_{H_2O}=0}$) in conjunction with observations from a previous study \cite{Fischer.2020}. Initially (I), there is a quick stress increase, quickly transitioning into a linear growth. In \autoref{fig:stress_dev}b, the deviation of the stress from the linear growth (point II) coincides with increasing divergence of the load from the constant rate period. Visual observation and profile measurements of the drying films reveal lateral drying \cite{Fischer.2020}. Due to an increased drying rate near the edges and pinning of the contact line between the coating and substrate, the film starts consolidating near the edges first, while the remainder of the coating is still supersaturated \cite{Chiu.1993b, Guo.1999, Holmes.2008, Holmes.2006, Ma.2005, Routh.1998, Fischer.2020}. This gradual increase in stress changes when a critical amount of menisci have formed, which leads to a more rapid increase in stress as the drying front propagates from the edges inward. When reaching point III (\autoref{fig:stress_dev}c), a very rapid stress increase towards the peak stress (IV) occurs. Images of the film show that in this period, the final supersaturated patch or region vanishes until a fully compacted film forms at the peak stress, leaving a pinhole or trench \cite{Fischer.2020}. Moreover, experiments with dyed heptanol suggest that the film is still completely saturated at that time \cite{Fischer.2020}. Following the peak stress, a two-step stress relaxation takes place (\autoref{fig:stress_dev}d). Stress release can have various superimposed causes. These causes range from the deformation and coalescence of soft particles \cite{Singh.2007}, to undesirable crack formation and plastic deformation, such as particle rearrangement \cite{Goehring.2013} often leading to unusable coatings. Another reason for stress relaxation is the reduction of stress inducing factors, i.e. evaporation of the menisci. When air starts invading the coating after point IV, the amount of liquid filled pores and menisci are reduced, leading to a decrease in equivalent pore stresses \cite{Coussy.2004, Price.2015}. We can exclude the formation of cracks since all examined films in this work were crack free. The reason for the two-step relaxation is not clear. However, a quick air invasion, i.e. menisci removal, along the edges of the entire cantilever could explain the fast decrease. Once the fractal drying front moves inward from all sides, the overall measured stress release decelerates due to the decreased area, and in particular, the shorter length where the stress acts. Eventually, the stress reaches a local minimum at point VI, before increasing again and approaching a final residual stress value (\autoref{fig:stress_dev}e). At the local minimum, the film is nearly dry, however, due to capillarity and blob formation, small amounts of the final residual fluid occupies the smallest pores \cite{Kharaghani.2013}. When air starts invading the remaining small liquid filled pores, the capillary pressure once again increases and, thus, the measured stress rises. 

The stress and load profile of a capillary suspension with $ \mathrm{\phi_{H_2O}/\phi_{solid}=0.125} $ is shown in \autoref{fig:stress_dev}f. After the start of the measurement (i), the stress increases with a transition to a linear growth (\autoref{fig:stress_dev}g). This transition occurs for the capillary suspension in a similar fashion as the pure suspension due to their nearly identical bulk fluid volumes. Unlike the pure suspension, when the stress increase deviates from its linear trend in the capillary suspension, the load still remains in the constant rate period. Despite the similar stress increase between point i and point ii, profile analysis shows significant differences between the two films. Initially, the capillary suspension also exhibits lateral drying, but to a lesser extent \cite{Fischer.2020}. The edges constantly dry but only partially consolidate while retaining this critical height. We attribute this phenomenon to a locally increased yield strength that is larger than the compressive forces, i.e. capillary pressure. Capillary suspensions have a large yield stress immediately after formulation, due to the presence of a sample spanning network \cite{Koos.2014}. In this type of pendular state suspension, the particles are connected by water bridges, forming flocs. Connections between these percolated flocs form a path (backbone) throughout the sample (sample spanning network) \cite{Bossler.2018}. As drying begins, the larger pores between the flocs compact (yield) and store some stresses. At the edges of the cantilever, where compaction has already occurred, the particle flocs form more percolating paths, which locally increases the yield stress and prevents the film from shrinking further, until the now higher intermediate yield strength is exceeded. While lateral evaporation persists, the denser packing near the edges wicks liquid from the surface supersaturated center of the cantilever until the yield strength of the film in the surface layers is equal across the cantilever. It is important to note that while the surface layers across the cantilever have compacted, the region below is still supersaturated. At point ii, this surface consolidation has finished and no supersaturated surface region is visible anymore \cite{Fischer.2020}. As drying proceeds, the capillary pressure increases beyond the intermediate yield strength, causing the film stress to grow much faster than its previously linear trend during further compaction. This rapid increase occurs much earlier for the capillary suspension than for the pure suspension ($ t_{ii}<t_{II} $). Qualitatively comparing the slope of the stress increase, we see a faster increase in the stress for the capillary suspension sample. A study by Price et al.~\cite{Price.2015}, where they used a walled cantilever to suppress edge drying, showed the effect of lateral drying on the stress evolution. While lateral drying causes a generally slower stress increase towards the peak, the absence thereof results in a very rapid stress growth. That is, the capillary suspension with its faster stress increase dries more uniformly across the cantilever without using artificial walls. Shrinkage profile measurements support this observation of top-down drying after point ii \cite{Fischer.2020}. While capillary suspension coatings exhibit larger superelevations near the edges of the cantilever compared to the regular suspension, this is due to coating application effects and are present already before drying. However, they retain their shape during the evaporation process in contrast to the regular suspension where particle accumulation (coffee-ring effect) is observed. The reason is the connected particle network and larger cluster structures compared to individually dispersed particles in the regular suspension. This also explains the almost uniform stress increase between point iii towards the peak stress (point iv), as depicted in \autoref{fig:stress_dev}h, which is further substantiated by the lack of drying defects such as pinholes or trench formation in the capillary suspension film. 

The point of maximum peak stress (iv) and, thus, the state of full film compaction displays several differences compared to the pure suspension. First, the peak is reached at an earlier time ($ t_{iv}<t_{IV} $). Second, the peak stress for the capillary suspension sample is lower ($ \sigma_{iv}<\sigma_{IV} $), which will be explored in more detail later on.  Moreover, while the constant drying rate for the pure suspension has already ceased at point II, the constant rate period for the capillary suspension persists even beyond full compaction. This implies that despite the early development of a consolidated surface, which usually decreases the drying rate, the drying continues unabated such that the surface must be kept sufficiently wet. We ascribe this phenomenon to corner flow \cite{Laurindo.1998, Prat.2011}, as described previously \cite{Fischer.2020}. After the peak stress (point iv), the capillary suspension also shows a two-step stress relaxation (\autoref{fig:stress_dev}i). In the capillary suspension, however, the first slope is clearly smaller in magnitude, i.e. displays a slower stress decay in the first step after the peak (-0.016~MPa/min) compared to the pure suspension (-0.022~MPa/min). Linking this first stress release to the drying rate and the fully compacted film, this implies capillary flow from inside the coating to the surface. This pore emptying through flow (decrease in equivalent pore pressure) must therefore be a slower process than pore emptying through Haines jumps. Additionally, we have shown in our previous work with dyed samples that point v corresponds to the appearance of the first dry areas transitioning to pore emptying by air invasion \cite{Fischer.2020}. The second stress release step is only marginally, but consistently slower for capillary suspensions (-0.010~MPa/min versus -0.011~MPa/min). This may be caused by the capillary water bridges, which require a larger air invasion pressure than the bulk fluid. Heptanol covers these bridges, which protects them, and prevents sudden stress releases by air invasion into pores through Haines jumps, as is the case for the pure suspension \cite{Blunt.2001, Scanziani.2018}. As a result, the heptanol cluster in the coating remains connected and should dry more gradually. Finally, the stress reaches its local minimum at point vi (\autoref{fig:stress_dev}j), before once again increasing towards the final residual stress. The width of the trough (full width at half maximum) is wider for capillary suspensions (6.45 min) than for the pure suspension (5.66 min), which could be explained by larger pores, formed by the initial network structure, filled with remaining fluid. This should theoretically also lead to a lower residual stress. This residual stress is very sensitive to film variations in terms of film height and lateral drying, which made it difficult to identify a clear trend of residual stress dependencies.

Capillary suspensions exhibit distinctly different stress development features compared to the pure suspension. For capillary suspensions, the stress increase is more rapid, indicating reduced lateral drying, whereas the stress decrease is more gradual. Moreover, the peak stress, which is an indicator for a film's proneness to cracking, is reduced in the capillary suspension. The differences and dependencies on drying conditions are examined more detailed in the following section.

\subsection{Peak stress comparison}

\begin{figure}[t!]
            	\capstart
            	\centering
            	\includegraphics[width=1\textwidth]{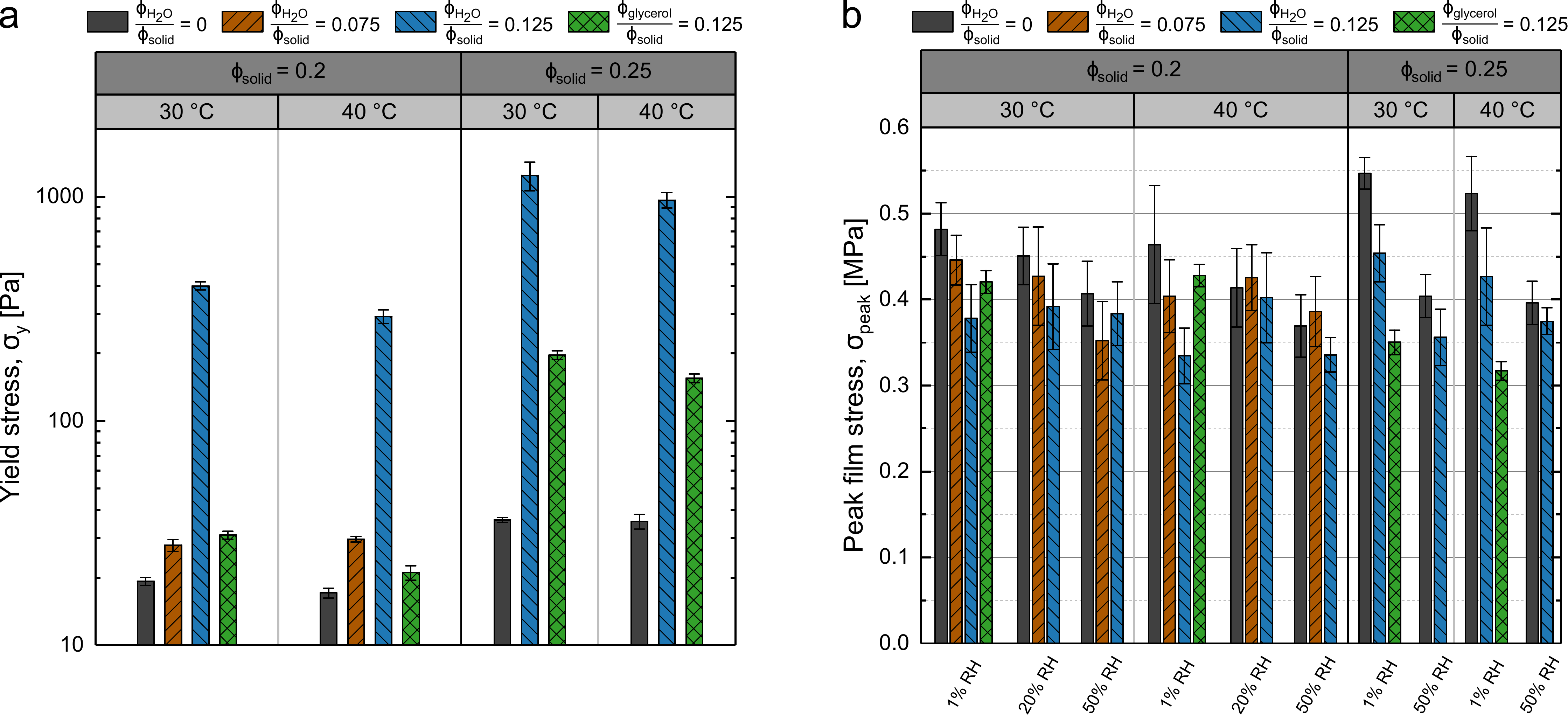}
            	\caption{(\textbf{a}) Yield stress measurements of samples with a variation in secondary fluid contents and initial particle volume fraction, as measured at 30 \degree C and 40 \degree C. (\textbf{b}) The comparison in peak film stress for drying experiments with a variation in secondary fluid content and initial particle volume fraction performed at 30 \degree C and 40 \degree C as well as different relative humidities.}
            	\label{fig:yield_peak_stress}
        \end{figure}

As observed in \autoref{fig:stress_dev}, the measured peak stress is lower for the capillary suspension with $ \mathrm{\phi_{H_2O}/\phi_{solid}=0.125} $ at $ \mathrm{\phi_{solid}= 0.2} $ and $ \mathrm{\phi_{solid}= 0.25} $ compared to the pure suspension. Generally, a sample without particle yielding exhibits larger stresses with increasing elastic strain acting on the sample network. The yield strength is, therefore, a measure of the coating's resistance to stresses. Because the yield stress of capillary suspensions can be modified through addition of water, the yield stress for the different formulations at the two temperatures is measured. The effect of a variation in secondary liquid volume on the yield stress, $ \sigma_y $ is marked with different colors (blue for the largest water addition) and patterns, as shown in \autoref{fig:yield_peak_stress}a. 
At a particle volume fraction of $ \mathrm{\phi_{solid}= 0.2} $, a relative volume fraction $ \mathrm{\phi_{H_2O}/{\phi_{solid}}= 0.075} $, only results in a small increase in yield stress. Further addition of water ($ \mathrm{\phi_{H_2O}}/{\phi_{solid}}= 0.125 $) causes a dramatic increase in the yield stress by more than one order of magnitude. At larger particle volume fraction ($ \mathrm{\phi_{solid}= 0.25} $), the thickening effect is even stronger. The addition of the same relative volume fraction of glycerol ($ \mathrm{\phi_{glycerol}/{\phi_{solid}}=0.125} $), on the other hand, barely increases the yield stress compared to the pure suspension sample for the lower particle loading. In contrast, the yield stress significantly increases by half an order of magnitude at $ \mathrm{\phi_{solid}= 0.25} $. An increase in temperature only has a small decreasing effect on the yield stress.

Naturally, the drying conditions will also influence the peak stress, as shown in \autoref{fig:yield_peak_stress}b. First, the drying temperature will influence the drying rate. At higher drying rates, one expects a larger peak stress due to smaller contact angles of the pinned menisci, leading to larger capillary pressures. A temperature induced decrease of the interfacial tension is not expected to balance the change in contact angle. Second, the relative humidity influences the drying of water while leaving the heptanol unchanged. The intention is to suppress drying of the capillary water bridges, such that the yield strength is maintained once the bridges are exposed to air. Exchanging water with the higher boiling point glycerol should further that effect. The influence of the discussed formulation changes and drying parameters on the maximum drying stress ($ \mathrm{\sigma_{peak}} $) is shown in \autoref{fig:yield_peak_stress}b. To understand these results, we will first only consider water as the secondary liquid and a particle volume fraction of $ \mathrm{\phi_{solid}=0.2} $. The black column (most left in each RH segment) represents the pure suspension. When evaporating into dry air (1\% RH), an increase in water fraction results in a lower peak stress, both at 30~\degree C and 40~\degree C. While an increase in humidity decreases the peak stress for no and low amounts of added water, it has no effect on the sample with the highest water addition (blue, $ \mathrm{\phi_{H_2O}}/{\phi_{solid}} = 0.125$), which is in contrast to our expectation. This could be explained by the fact that the water capillary bridges remain until the stress inducing meniscus has receded further into the coating. Interestingly, the highest relative humidity (50\% RH) significantly reduces the peak stress of all water fractions to a level comparable with the high yield strength capillary suspension ($ \mathrm{\phi_{H_2O}/\phi_{solid}=0.125} $) at both examined temperatures. At elevated temperatures, one would expect larger peak stresses due to a reduction in the heptanol contact angle caused by the pinned menisci. Instead, there is no influence of temperature on the peak stress for the pure suspension. For the $ \mathrm{\phi_{H_2O}/\phi_{solid}=0.125} $ and $ \mathrm{\phi_{solid}=0.2} $ capillary suspension sample, the peak stress tends to decrease at higher temperature and 1\% RH (p $ \le 0.07 $, i.e. the means are different with a probability of more than 93 percent). From the small yield stress decrease (\autoref{fig:yield_peak_stress}a), one would have expected a larger peak stress in the absence of particle migration. However, the bulk viscosity of heptanol is also temperature dependent. It decreases from 5.090 $ \mathrm{mPa\cdot s} $ at 30 \degree C to 3.740 $\mathrm{ mPa\cdot s} $ at 40 \degree C \cite{EstradaBaltazar.2015}. This reduction may allow particle clusters that are not part of the sample spanning network, or have yielded already, to rearrange more easily resulting in a partial stress release. This rearrangement would most likely occur in the lower layers, which are still uncompacted before the stress peak.

The pure samples show more pronounced lateral drying at higher temperatures \cite{Fischer.2020}. This more pronounced lateral drying leads to trench formation and causes more particles to migrate, thus releasing the evolving stresses. Therefore, the maximum stress, which is expected to be larger compared to a lower temperature cannot be independently measured since the stress accumulation and stress relaxation through yielding and particle migration are superimposed. Increasing the initial solid load to $ \mathrm{\phi_{solid}= 0.25} $ leads to smaller liquid volume between the particles. In non-stabilized suspensions, this leads to more particle-particle interactions upon lateral drying, with the result of restricted particle migration (reduced trench formation and pinholes) and, consequently, a larger stress formation due to the absence of stress relaxation through particle migration \cite{Guo.1999}. Indeed, as shown in \autoref{fig:yield_peak_stress}b, the pure suspension at 1\%~RH displays a larger peak stress than at lower volume fraction. This trend holds for both drying temperatures. Interestingly, the capillary suspension $ (\mathrm{\phi_{H_2O}/\phi_{solid}=0.125}) $ peak stress at 1\%~RH is also larger for the larger particle load. Increasing the temperature does not significantly affect the peak stress. Due to the larger particle-particle interactions at $ \mathrm{\phi_{solid}=0.25} $ than at the lower solid load, the sample-spanning network is reinforced and particle rearrangements are inhibited. Increasing the relative humidity at higher particle loads again significantly lowers the peak stress of the pure suspensions. Even for the capillary suspension, there is an additional decrease in the peak stress with increasing humidity that is not observed in the lower particle volume fraction.

Exchanging the secondary fluid for glycerol has two immediate effects. First, the yield stress for a capillary suspension with the same volume of secondary fluid $ \mathrm{\phi_{glycerol}/}{\phi_{solid}}=0.125 $ is much lower than when using water, especially at $ \mathrm{\phi_{solid}= 0.2} $ (\autoref{fig:yield_peak_stress}a). This change is due to the lower interfacial tension of glycerol. Secondly, glycerol evaporates at 290 \degree C, which means it essentially does not evaporate under the examined conditions and should remain completely in the film. As expected due to the only slightly larger yield stress for the glycerol capillary suspension at lower particle load, the reduction in peak stress compared to the pure suspension is small (\autoref{fig:yield_peak_stress}b). In contrast, the larger $ \sigma_y $ at higher initial solid content causes an immense reduction in peak stress of approximately 40\%. This exceeds the reduction measured for the water sample at 1\% RH despite the higher yield stress. This observation, along with the decrease in peak stress at 50\% RH, suggests that water lowers the coating's resilience against shrinking stresses. 

We have shown that the peak stress can be reduced by using capillary suspensions in drying coatings. Changing the drying temperature did not generally lead to a change in the peak stress. Interestingly, we found that an increase in humidity can lower the peak stress of the pure suspension and capillary suspension at high particle load (whereas no change is observed at $ \mathrm{\phi_{solid}=0.2} $). Capillary suspensions change the microstructure of the final film. Since larger pores should reduce the capillary pressure and thus the peak stress during drying, we further examine these effects in the following section.

\subsection{Film morphology at the peak: Residual fluid}

\begin{figure}[t!]
            	\capstart
            	\centering
            	\includegraphics[width=0.5\textwidth]{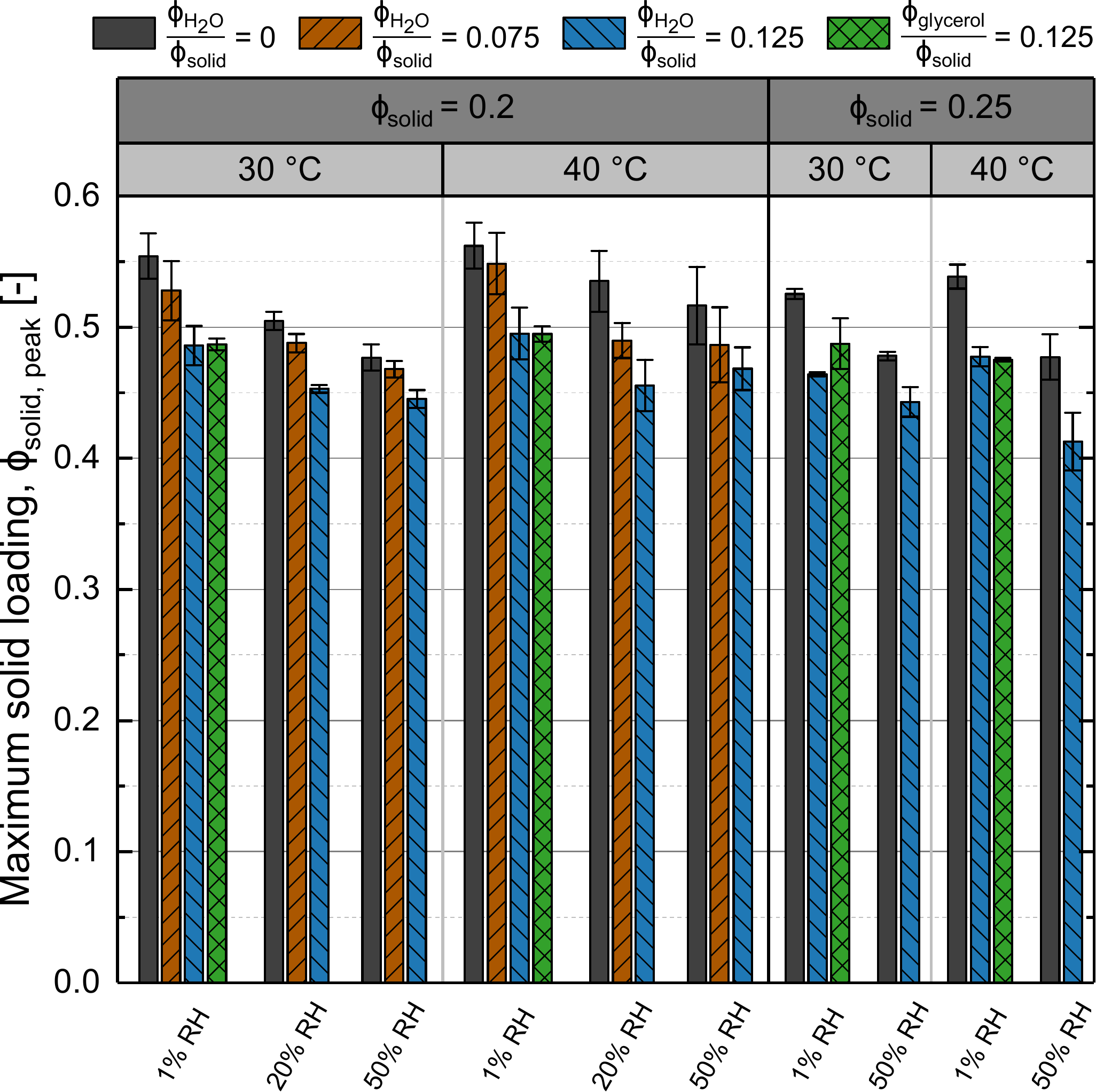}
            	\caption{The average dry film particle volume fraction of the experiments obtained from the particle volume fraction at the peaks for the aforementioned formulations and drying conditions.} 
            	\label{fig:PVF}	
        \end{figure}

In the previous section, we observed the addition of a secondary fluid, which changes their yield strength, decreases the peak stress. While non-stabilized suspensions can flocculate due to van der Waals forces, a capillary suspension represents a highly flocculated system where the capillary force (originating from the secondary fluid bridges), is much stronger \cite{Koos.2011, Koos.2014}, leading to the yield stress increase as displayed in \autoref{fig:yield_peak_stress}a. However, this network formation may also result in a different microstructure of the coating. The film density, that is the final packing volume of a coating, is a measure of its porosity. The simultaneous stress and weight measurement used in this study allows us to directly estimate the final packing volume fraction in a reliable way based on a few simple assumptions:
        \begin{enumerate}
            	\item \label{itm:gly_vol} For the systems with added glycerol, the glycerol is treated as inert and remaining completely in the film. 
            	\item \label{itm:H2O_vol} For the systems with added water, the water volume in the capillary bridges is treated as heptanol (the densities are approximately equal and there is no difference in their distribution within the film).
            	\item \label{itm:dry} When the stress has approached its residual value, the film was considered dry (devoid of any volatile components). Due to small noise in the weight signal, the final dry mass $ \mathrm{m_\mathrm{final}} $ is the averaged value over the last 100 recorded data points (50 seconds).
            	\item \label{itm:liq_cont} The volatile liquid content of the dry film (isothermal sorption equilibrium) is negligible ($\mathrm{V_\mathrm{final} = V_\mathrm{Al_2O_3}} $ or $\mathrm{V_\mathrm{final} = V_\mathrm{Al_2O_3} + V_\mathrm{glycerol}} $).  
            	\item \label{itm:sat} The coating is fully saturated and the maximum final packing volume fraction is obtained when the peak stress $ \mathrm{\sigma_{peak}} $ is reached \cite{Guo.1999}.
        \end{enumerate}
The experiments were carried out until the residual stress approached its final value (assumption~\ref{itm:dry}). This is sufficient since the weight change was already below the small fluctuations of the balance a few minutes after point VI/vi. Glycerol with a boiling point of 290~\degree C will not evaporate in measurable amounts at 40~\degree C (assumption~\ref{itm:gly_vol}). For the cases with added water, small amounts of water may remain in the final film. The error induced by assumption~\ref{itm:H2O_vol} is below 0.6\% due to the low secondary fluid volume fractions used in capillary suspensions. In a previous study with dyed 1-heptanol, we found that the film is saturated until the peak stress had occurred (assumption~\ref{itm:sat}) \cite{Fischer.2020}. 
These assumptions allow to calculate the final packing volume fraction $ \mathrm{\phi_{solid,peak}} $ as follows:
        \begin{align}
                \mathrm{Water:} \quad \phi_\mathrm{solid,peak}&= \frac{V_\mathrm{Al_2O_3} 
                }{V_\mathrm{Al_2O_3} + V_\mathrm{heptanol}}  
                        = \frac{\frac{m_\mathrm{final}}{\rho_\mathrm{Al_2O_3}}}{\frac{m_\mathrm{final}}{\rho_\mathrm{Al_2O_3}} + \frac{m(\sigma_\mathrm{peak}) - m_\mathrm{final}}{\rho_\mathrm{heptanol}}} \label{eq:pvf} \\
                \mathrm{Glycerol:} \quad \phi_\mathrm{solid,peak}&= \frac{V_\mathrm{Al_2O_3} 
                }{V_\mathrm{Al_2O_3}+V_\mathrm{glycerol} + V_\mathrm{heptanol}}  
                        = \frac{\frac{m_\mathrm{final}}{\rho_\mathrm{Al_2O_3}+0.125\rho_\mathrm{glycerol}}
                        }
                        {\frac{m_\mathrm{final}}{\rho_\mathrm{Al_2O_3}+0.125\rho_\mathrm{glycerol}}(1+0.125)+ \frac{m(\sigma_\mathrm{peak}) - m_\mathrm{final}}{\rho_\mathrm{heptanol}}} 
                \label{eq:pvf_gly}
        \end{align}
with $ m(\sigma_\mathrm{peak}) $ indicating the mass at the stress peak and $ \phi_\mathrm{solid,peak} $ denoting the final volume packing. \autoref{eq:pvf} is used for the samples with water, and \autoref{eq:pvf_gly} shows the equation for the samples with glycerol. The factor of 0.125 arises from the initial volume fraction $ \phi_\mathrm{glycerol} $/$ \phi_\mathrm{Al_2O_3} = 0.125$ in these samples. The calculated particle volume fraction at the stress peak, $ \mathrm{\phi_{solid,peak}} $ of our experiments is shown in \autoref{fig:PVF}. 
In general, we see a decrease in particle volume fraction at the stress peak $ \mathrm{\phi_{solid,peak}} $ with increasing water content. That is, inducing a flocculated particle network increases the porosity of the sample. As can be seen in \autoref{eq:p_cap}, an increase in pore size results in lower capillary stresses, which in turn leads to a lower peak stress for the same film height, or in other words, a higher critical cracking thickness. This observation is in accordance with the findings of Guo and Lewis~\cite{Guo.1999}, and Singh et al.~\cite{Singh.2009}. Singh et al. found that for a constant particle loading for alumina suspensions dispersed in water, the final packing volume of the coating decreased with increasing degree of flocculation, i.e. the film contained more pores \cite{Singh.2009}. Similarly, Guo and Lewis found the same decrease in dry film volume fraction for flocculation induced by salt addition to stabilized silica particles \cite{Guo.1999}. Additionally, their experiments indicated that an increase in initial particle volume fraction led to a larger dry film volume fraction, i.e. a more compact coating.

Increasing the relative humidity tends to slightly decrease the dry film particle volume fraction in the capillary suspensions. Interestingly, an increase in relative humidity also increases the dry film porosity (decreases $ \mathrm{\phi_{solid,peak}} $) of coatings prepared from the pure suspension, coatings that were not intentionally flocculated. This effect appears to be more evident at lower drying temperature. A temperature increase at the initial particle volume fraction of 0.2 results in a denser film at each $ \mathrm{\phi_{sec}} $ and relative humidity. This finding supports our hypothesis of larger particle and cluster mobility at 40~\degree C, which allows some stress relaxation through particle migration. Increasing the initial particle load to 0.25 causes a less dense film for both temperatures compared to the lower initial load, a result that is in accordance with the larger measured yield stress (higher degree of flocculation). The particle interactions suppress the particle mobility resulting in more porous films. This result is in contrast to the results of Guo and Lewis~\cite{Guo.1999}. The discrepancy may arise from the use of partially stabilized silica particles versus unstabilized alumina, as well as the use here of irregularly shaped alumina particles.

The link between the film porosity and stress peak should be established for capillary suspensions in order to elucidate the stress reduction potential caused by the secondary capillary bridges rather than through a porosity increase. The samples with added water both show a decrease in $ \mathrm{\sigma_{peak}} $ and $ \mathrm{\phi_{solid,peak}} $, as predicted. Recall that at lower initial solid load, there was no difference in peak stress with the capillary suspension ($ \mathrm{\phi_{H_2O}/\phi_{solid}=0.125} $) and increasing relative humidity at either temperature (\autoref{fig:yield_peak_stress}b). Yet, when dried at 50\% RH, the dry capillary suspension films are more porous (\autoref{fig:PVF}). This is a first indication that a porosity increase does not significantly contribute to a peak stress reduction in capillary suspensions. The glycerol capillary suspension samples at 1\% RH and both temperatures provide a clear answer. Both capillary suspensions with $ \mathrm{\phi_{sec}/{\phi_{solid}}=0.125} $ show almost identical porosities, while they differ in peak stress. This clearly demonstrates that particle-particle interactions can dominate the influence on the peak stress as opposed to the porosity of the coating. 

In this section, we have shown that capillary suspensions lead to more porous dry films than the pure suspension due to flocculation. However, increasing the relative humidity can also increase the porosity, in particular for the pure suspension. Moreover, we have demonstrated that the porosity increase in capillary suspensions are not responsible for the reduction in peak stress. Instead, the particle interactions caused by the secondary liquid capillary bridge determine the influence on the peak stress. This suggests that by choosing a secondary liquid with a higher boiling point than water and stronger interfacial tension than glycerol can further improve the formulation.   

\subsection{Residual saturation}
\label{sec:Residual}

\begin{figure}[t!]
            	\capstart
            	\centering
            	\includegraphics[width=0.5\textwidth]{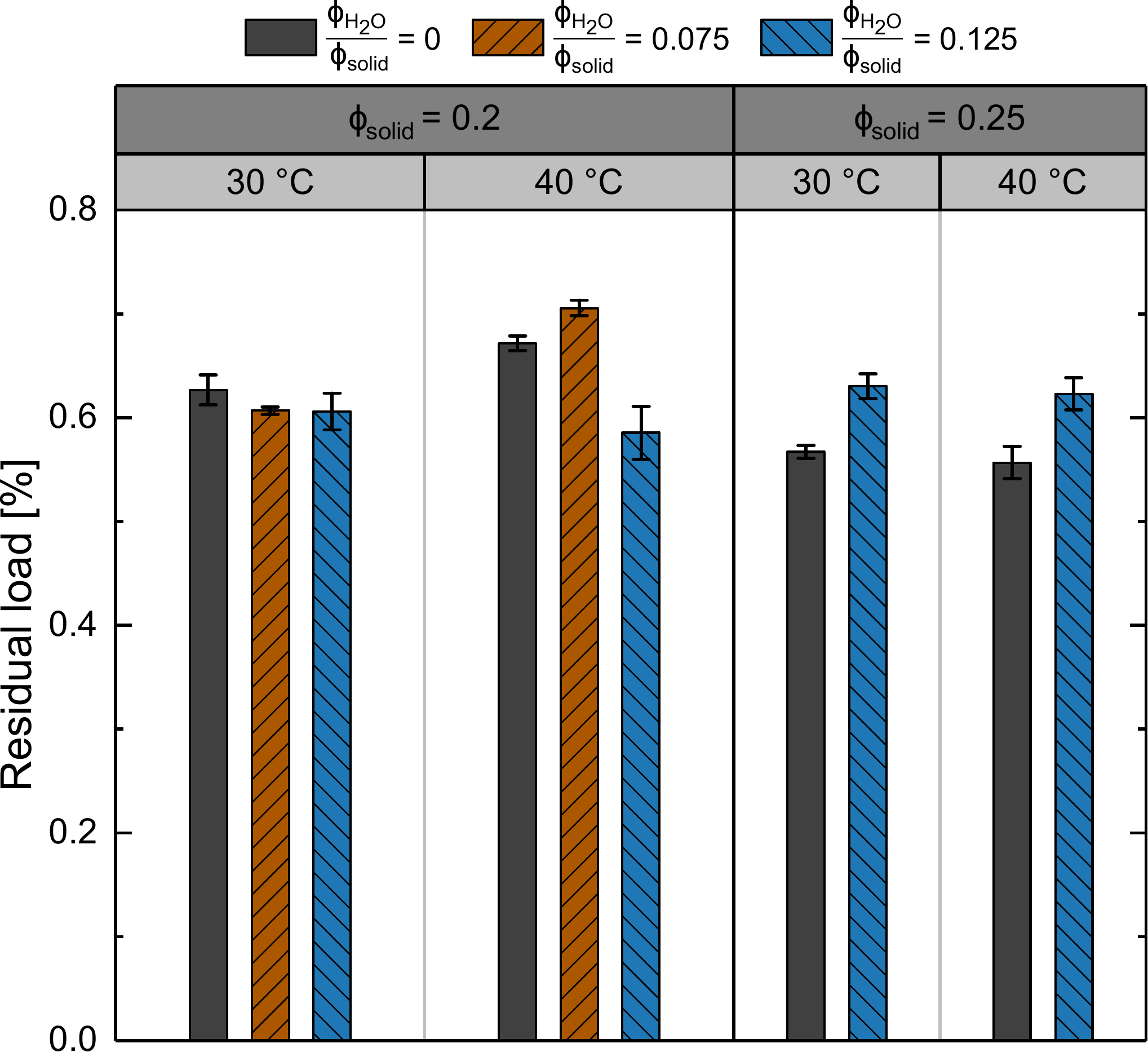}
            	\caption{Residual fluid content for differently formulated suspensions with increasing water content and initial particle load. The samples were dried at a constant temperature of 30 \degree C and 40 \degree C. Subsequent mass loss caused by heating was recorded with a thermogravimetric analyzer. The residual liquid load is normalized against the dry particle weight.} 
            	\label{fig:Residual}	
        \end{figure}

The observed temperature dependence of the secondary liquid on the dry film porosity raises the question whether the capillary water bridges persist until after drying or if they evaporate earlier, as would be predicted by the higher vapor pressure. If the water bridges partially remain in the film, one would expect a larger residual fluid content and longer influence of the bridges on the drying dynamics. In order to detect these relatively low residual contents, a thermogravimetric analyzer with a resolution of 0.1 \micro g was used. The samples were dried into a dry nitrogen atmosphere and a similar wet film thickness in the crucible as on the cantilever, as described in \autoref{sec:TGA}. \autoref{fig:Residual} 
shows the residual liquid in terms of the coating's load $ m_\mathrm{liquid}/m_\mathrm{dry} $ (\autoref{eq:load}) for films prepared with increasing water content at different initial solids load for both 30~\degree C and 40~\degree C. All experiments show a residual load lower than the initial water load when formulating the samples (3.2 \% for $ \mathrm{\phi_{H_2O}/\phi_{solid}=0.125} $ at $ \mathrm{\phi_{solid}=0.2} $), demonstrating that at least 80\% of the water has evaporated.  At an initial solid fraction of $ \mathrm{\phi_{solid}=0.2} $, there is no difference in residual load for capillary suspensions when dried at 30~\degree C. At the higher temperature, both the pure suspension and weak capillary suspension show an increase in the residual fluid fraction, while the capillary suspension exhibits a lower saturation compared to 30~\degree C. At higher initial solid fraction, the results contrast to the lower particle fraction; while the residual liquid content for the pure suspension is at a lower level as for $ \mathrm{\phi_{solid}= 0.2} $, the capillary suspension displays a clear increase. This means that more residual fluid, presumably water, remains in the capillary suspension at $ \mathrm{\phi_{solid}= 0.25} $.

Comparing the residual liquid content (\autoref{fig:Residual}) to measurements of the film porosity at 1\% RH (\autoref{fig:PVF}), we see that the film porosity increases with increasing water content, i.e. a stronger gel-network and thus more flocculated suspension. This in turn could lower diffusional resistances for vapor transport from within the coating to the outer surface, leading to a lower film saturation. The residual fluid load of the pure suspension closely correlates with the film porosity. An increase in temperature increases the diffusivity, which further enhances mass transport to the surface. However, the results for $ \mathrm{\phi_{solid}=0.25} $ are in contrast to this theory. While the porosity of the capillary suspension films at higher initial solid fraction are somewhat higher, the retained amount of liquid also increased. Further insight can be provided by research using porous media, especially in oil recovery, where three phase flow phenomena are more common. When modeling the drying of a porous system, the rules of invasion percolation (IP) and drainage flow can be applied \cite{Laurindo.1998, Segura.2005}. Drying of the pure suspensions leads to phenomenon such as Haines jumps or meniscus snap-off (more dominant at imbibition conditions) \cite{Segura.2005, JoekarNiasar.2012}. These incidents lead to isolated blobs of heptanol that are disconnected from the bulk. These droplets can remain trapped in the film as residual fluid. As mentioned in \autoref{sec:stress} and shown in \autoref{fig:stress_dev}, the water bridges, with a higher capillary pressure than heptanol, allows the capillary suspension film to dry at a constant rate even beyond close packing. The capillary bridges formed by the secondary liquid obstruct the bulk fluid pathways so that snap-off events are reduced or even prevented. Furthermore, heptanol covers the water bridges, even in the presence of air \cite{Blunt.2001}. Thus, the bulk heptanol remains connected to a larger degree, leading to piston-like flow \cite{JoekarNiasar.2012}. Continued drainage (drying) can even lead to reconnections between previously disconnected oil clusters \cite{Armstrong.2012}. In a recent micro-CT study by Scanziani et al., drainage experiments with brine, oil, and gas showed that water is retained in the smallest necks and pores upon gas invasion in a porous carbonate rock \cite{Scanziani.2018}. Oil occupies the medium sized pores and gas preferentially invades the larger pores. In porous media filled with oil and water, an increase in temperature additionally favors water retention leading to preferential oil drainage \cite{Davis.1994, She.1998}.

With this more complex behavior in mind, we can revisit the potential difference in structural particle arrangements for the capillary suspension at $ \mathrm{\phi_{solid}=0.25} $ compared to $\mathrm{\phi_{solid}=0.2} $. For granular materials, the coordination number $ z $ should be related to the particle volume fraction by $ z=\pi / (1-\phi_\mathrm{solid}) $ \cite{Pietsch.1967}. Since the number and volume of capillary bridges are related through $ \mathrm{\phi_{sec}/\phi_{solid}= z/2 \cdot V_{bridge}/V_{particle}} $ the bridge volume must be similar for the higher particle fraction sample since the $ \phi_\mathrm{sec}/ \phi_\mathrm{\phi_{solid}} $ ratio remains constant \cite{Heidlebaugh.2014,Koos.2014}. Of course, capillary suspensions are composed of dense flocs connected by a sparse backbone \cite{Bossler.2018} where the porosity of these networks is only weakly related to the particle fraction \cite{Dittmann.2012}. In accordance with the porous rock/ particle bed findings, this would lead to more residual liquid stemming from the similar capillary water bridges. The increase in the number of connections would explain the observed increase in yield stress, smaller peak stresses, similar porosity, and larger residual fluid content compared to the respective capillary suspension sample at lower initial particle load.

We conclude that the residual fluid content for the pure suspension is mainly dependent on the porosity of the final dry film. For the capillary suspension on the other hand, it appears to be an interplay of porosity and number of capillary water bridges. Blockage of pathways for air invasion leads to better drying of heptanol even fortified at elevated temperatures. This suggests that residual liquid in capillary suspensions is predominantly water. The extent should be controlled by the total amount of water bridges present in the sample. The following section will therefore examine if capillary water bridges despite its higher vapor pressure persist longer than heptanol. 

\subsection{Capillary bridge evaporation}

\begin{figure*}[t!]
            	\capstart
            	\centering
            	\includegraphics[width=0.75\textwidth]{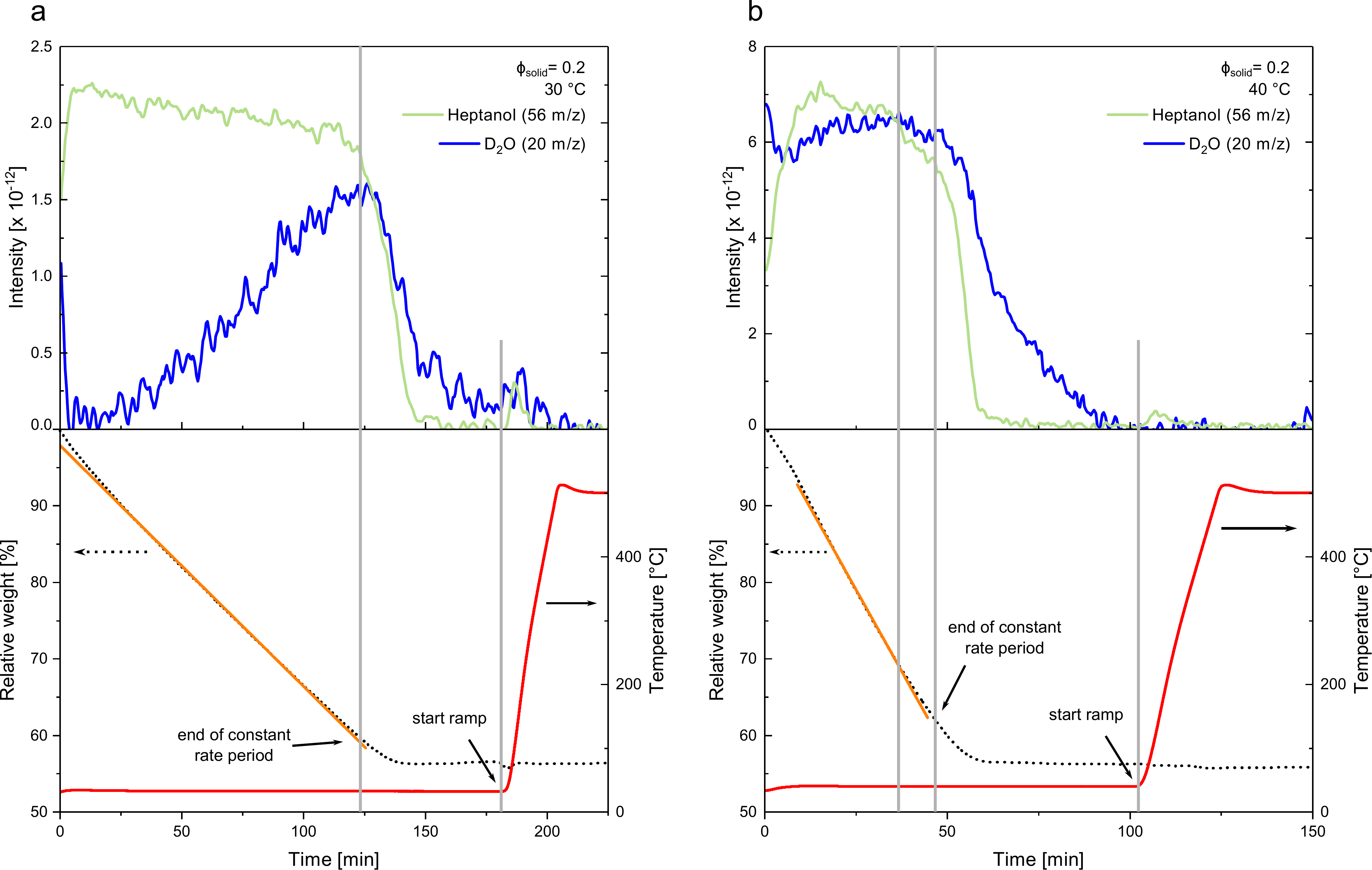}
            	\caption{Results for thermogravimetrical analysis coupled with mass spectrometry over time for drying of capillary suspensions with initial volumetric particle load of 0.2 and dried at (\textbf{a}) 30 \degree C and (\textbf{b}) 40 \degree C. The upper panel illustrates the intensity of the heptanol measured at a specific mass to charge ratio of 56 m/z (green line) and the intensity of the heavy water mass at 20 m/z (blue line). The lower panel shows the temperature profile (red line) and the relative weight change of the coating (dotted black line).}
            	\label{fig:MS}
        \end{figure*}

In the previous section, we put forward a hypothesis that the water capillary bridges persist through the late stages of drying. In order to directly elucidate the evaporation behavior of the capillary water bridges, we coupled a mass spectrometer (MS) with thermogravimetrical analysis (TGA), as shown in \autoref{fig:MS}. 
This combination allows us to set an accurate temperature, while obtaining the drying rate and analyzing the headspace composition. The temperature for these measurements followed the profile shown in \autoref{tab:TGA} where the sample was first dried at either 30 \celsius~ (\autoref{fig:MS}a) or 40 \celsius~ (\autoref{fig:MS}b) then, once the sample mass remained constant, increased to 500 \celsius. Since 1-heptanol (56 m/z) also has a fractional mass peak overlapping with water, we replaced water with D$_2$O (20 m/z). Despite flushing, we observed a constant decrease for the 20 m/z signal, the indicator for heavy water, even during the temperature increase to 500~\degree C. As this is highly unlikely to be D$_2$O, we hypothesize that the argon, which is also present in trace amounts and also possesses a fractional peak at 20 m/z, could be the reason for the constant decline in that signal. Therefore, we used the linear decrease after attaining a temperature of 500 \degree C as a baseline for correction in the D$_2$O signal. Subsequently, the MS signals were smoothed. \autoref{fig:MS} shows the results for capillary suspensions at an initial solid load of 0.2 and $ \mathrm{\phi_{H_2O}}/{\phi_{solid}} = 0.125$ at the two temperatures. The green solid line represents the 56 m/z intensity over time for heptanol and the blue solid line shows the intensity of the 20 m/z signal detected for heavy water. After the start of the experiment for drying of the capillary suspension at 30~\degree C (\autoref{fig:MS}a), 1-heptanol constantly evaporates until the end of the constant rate period denoting the transition to the second drying period in which the headspace concentration drastically drops. When the film appears gravimetrically dry, the heptanol in the gas phase swiftly decreases. Once the temperature is rapidly increased, the small amount of residual heptanol evaporates. 

The evaporation of heavy water from the film is quite different. Over the course of the constant rate period, the gas phase concentration of D$_2$O continually increases. In fact, it is still increasing at the end of the constant rate period when the heptanol concentration starts to decrease. Once the heavy water concentration starts to decline, the rate of decrease is smaller than for the heptanol. Furthermore, water is still found in the gas phase well after the film appears to be dry (mass reaches a constant value) and the temperature increase evaporates the leftover water. This clearly shows that at 30~\degree C, the capillary bridges dry much slower than heptanol despite the higher bulk vapor pressure. More importantly, the capillary bridges persist and are still drying when heptanol has already reached its equilibrium. 

The drying of the capillary suspension at 40~\degree C is shown in \autoref{fig:MS}b. As with the drying at 30~\degree C, heptanol dries with a fairly constant rate, although there is more gradual decrease before the end of the constant rate period. Due to the high sensitivity of the MS, the reason for this earlier drop could be inhomogeneities in coating thickness. This is also reflected in the TGA measurement where a rate change is observed in the film weight. Nevertheless, a sharp decrease is observed after entering the second drying period. Again, the water concentration drop is delayed by several minutes, once again occurring when the film has nearly reached its dry state. Surprisingly, this delay is longer at the higher temperature, supporting the results obtained from drainage experiments \cite{Davis.1994, She.1998} . When the heptanol concentration in the gas phase is approximately zero, the water still has a high concentration in the gas phase. This concentration decays slowly, only reaching zero just before the temperature increase. Upon heating, the trapped heptanol evaporates, but water is not detected. This suggests that while persisting longer at elevated temperature, the water has completely evaporated before the temperature rise. 

These results show that the capillary bridges, despite being formed of water with a much lower boiling point than 1-heptanol, persist for long times during drying. While the cause for this delay should be determined, it does prove that even secondary fluids with a higher vapor pressure can persist in capillary suspension networks. The full potential will only be realized by replacing water with higher boiling point wetting liquids, such as glycerol, to delay or even prevent bridge evaporation. The lower interfacial tension of glycerol, however, tempers this potential. Thus, liquid combinations with high interfacial tension in addition to the desired high boiling point must be sought. One way of tailoring the heptanol-water system used in this study is by adding salt to the water. This decreases the vapor pressure and even increases the interfacial tension at the same time \cite{Zhang.2012}.

The present research also points to a few unanswered questions. First, why a difference between the evaporation dynamics is observed in the D$_2$O between 30~\degree C and 40~\degree C (\autoref{fig:MS})? While the rate of the evaporation for the heptanol remains unchanged, the evaporation rate for the D$_2$O appears to be constant in time at the higher temperature whereas it increases with time at the lowest temperature. We hypothesize that the difference may arise due to the small, but finite solubility of the D$_2$O in the heptanol. Thus, the change may be due to a varying miscibility of D$_2$O in heptanol. While the air-liquid  interface  at  the  surface  may  be  depleted  of D$_2$O at 30~\degree C , the D$_2$O at the air-liquid interface is being constantly and more rapidly replenished from the bulk liquid at 40~\degree C. This is caused by the larger solubility and thus concentration close to the capillary bridges, leading to a larger gradient. Additionally, mass transport is increased due the higher temperature. The influence of particle size on the present results is also of interest. Previous research has shown that the yield stress depends on the reciprocal radius \cite{Koos.2012}. The structure of these systems, however, will differ networks created with larger particles having a higher fractal dimension \cite{Bossler.2018}. This also changes the average size of the flocs (relative to the particle size) and the strength of the inter-floc links. The particle size also has clear implication to the drying of these suspensions. The particle size will change the capillary pressure (again, with a reciprocal dependence). Would this change be balanced by the change in the yield stress? There is also a change in the particle (or floc) mobility. Thus, the dependence of the particle size on the stress and structure of these drying capillary suspensions may be very interesting for further research. 

\section{Conclusions}
\label{conclusions}
 
In this paper, we investigated effects accompanying the drying of suspensions with and without capillary interactions. Enhancing a pure, oil-based suspension with only a few drops of water induces a network, transforming the suspension into an elastic paste. During drying, stresses caused by the capillary pressure within pores arise that are opposed by the water bridges in the capillary suspensions. Therefore, we simultaneously measured the evolving stresses and compared them with the drying rate. The capillary suspensions exhibit a faster stress rise, an indication of more uniform drying, and have lower peak and residual stresses. The peak stress decreases with increasing amounts of water, such that films formed from capillary suspensions are less prone to cracking. Additionally, we found that drying into more humid air enhances the stress reduction, even for the pure suspension. An increased film porosity, caused by capillary suspension networks, are only minor contributions to the observed stress reduction. Moreover, with reduced particle migration during drying, for example at higher initial solid loads, capillary suspensions show a greater potential. Lastly, we have shown that the capillary water bridges, despite having a lower bulk boiling point, persist into late stages of drying, after the heptanol has mostly evaporated. The partial evaporation of capillary bridges during drying is different at the two temperatures examined, which should be investigated further as well as the reason for the persistence of the bridges. Tuning the suspension with a higher boiling point secondary liquid can further this potential. In summary, the drying of capillary suspensions is a complex interplay between three phase flow, interfacial tension, capillary pressure dependencies, vapor pressure and the resulting yield strength. Understanding this interplay will help us tune the drying behavior for the desired applications.

\section*{Acknowledgements}

The authors would like to thank Almatis GmbH for the donation of alumina particles. Additionally, we thank Prof. Rob Ameloot and Dr. Min Tu for access and help with the mass spectrometry measurements at the KU Leuven Centre for Membrane Separations, Adsorption, Catalysis, and Spectroscopy for Sustainable Solutions (cMACS), Belgium. Finally, we acknowledge financial support from the German Research Foundation, DFG under project number KO 4805/2-1 and the Research Foundation Flanders (FWO) Odysseus Program (grant agreement no. G0H9518N).

\section*{Conflicts of interest}
The authors declare no conflicts of interests



\end{document}